\documentclass{PoS}
\usepackage{amsmath,amssymb,bm}
\usepackage{array}

\allowdisplaybreaks[3]
\title{Double parton scattering for perturbative transverse momenta}

\ShortTitle{Double parton scattering for perturbative transverse momenta}

\author{\speaker{Maarten G.A. Buffing}\\%
	Deutsches Elektronen-Synchrotron DESY\\
	22603 Hamburg, Germany\\
	E-mail: \email{maarten.buffing@desy.de}}

\author{Markus Diehl\\
	Deutsches Elektronen-Synchrotron DESY\\
	22603 Hamburg, Germany\\
	E-mail: \email{markus.diehl@desy.de}}

\author{Tomas Kasemets\\
	Nikhef and Department of Physics and Astronomy, VU University Amsterdam\\
	De Boelelaan 1081, NL-1081 HV Amsterdam, the Netherlands\\
	E-mail: \email{kasemets@nikhef.nl}}

\abstract{The cross section for transverse momentum dependent double parton scattering involves transverse momentum dependent double parton distributions (DTMDs). In the region of perturbative transverse momentum the DTMDs can be matched onto collinear double parton distributions. We present the framework and results for this matching, as well as the evolution equations for DTMDs in the region of large distance between the two partons. We discuss explicit results for one-loop matching coefficients and evolution kernels.}

\FullConference{ QCD Evolution 2016\\
				May 30-June 03, 2016\\
				National Institute for Subatomic Physics (Nikhef) in Amsterdam}

\begin{document}

\section{Introduction}
\label{ss:introduction}
In the collision of two hadrons, double parton scattering (DPS) describes interactions in the form of two hard processes, each initiated by a separate set of partons. DPS was already considered long ago~\cite{Landshoff:1975eb,Landshoff:1978fq} and its understanding is relevant for studying physics at particle colliders such as the LHC~\cite{Bartalini:2011jp,Proceedings:2016tff}. In these proceedings we will focus on DPS with color singlet final states. An example of such a process is e.g.~double Drell-Yan (DY), for which a factorization formula was first written down in the Refs.~\cite{Paver:1982yp,Mekhfi:1983az}. 

\begin{figure}[!b]
\begin{center}
\includegraphics[width=0.5\textwidth]{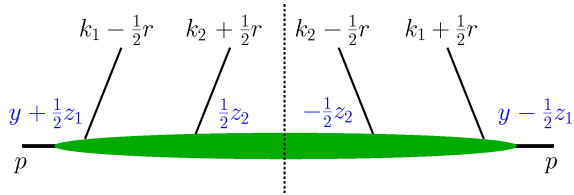}
\caption{The momenta involved in double parton scattering illustrated for a double parton correlator. The momentum space configuration involves the momenta $k_{i}$ and $r$, whereas $\textcolor{blue}{z_{i}}$ and $\textcolor{blue}{y}$ (blue color online) are the configuration space variables.}
\label{f:labeling}
\end{center}
\end{figure}

In DPS processes, correlators are described by double parton distribution functions (DPDFs) and double transverse momentum dependent distributions (DTMDs). In configuration space the correlator is described by a combination of the parameters $z_1$, $z_2$ and $y$, see Fig.~\ref{f:labeling} for an illustration. The transverse distance $\bm{y}$ is a measure for the separation between the two hard processes~\cite{Paver:1982yp,Mekhfi:1983az,Diehl:2011yj}. In momentum space the situation is described by a combination of the momenta $k_{1}$, $k_2$ and $r$. For the discussion in these proceedings, we will consider the short-distance expansion, where the two hard processes have a large spatial separation and where both transverse momenta $\bm{k}_1$ and $\bm{k}_2$ are perturbative. As such, $\bm{z}_1$ and $\bm{z}_2$ are small compared with a nonperturbative scale $1/\Lambda$ and $\bm{y}$ is of the order $1/\Lambda$. Although we use these approximations, many results we present are valid beyond it.

In these proceedings, we will give details of factorizing the DPS soft factor into $z_{1}$, $z_{1}$ and $y$-dependent parts. An evolution kernel $K$, familiar from the context of single parton scattering (SPS)~\cite{Collins:1981uk,Collins:1981uw,Collins:1984kg,Collins:2011zzd}, is related to this soft factor. Furthermore, one can choose the two hard processes to have separate renormalization scales $\mu_{1}$ and $\mu_{2}$. There is a scale $\zeta$ on top that plays the role of a rapidity regularization scale. We will solve their corresponding evolution equations. Finally, we provide the matching equations of DTMDs onto DPDFs for processes with a colorless final state, which is the main goal of these proceedings. For more details we refer to a forthcoming paper~\cite{Buffing:2016}.

\section{Soft factors and color}
\label{ss:soft_factors}
In DPS not only the two hard interactions have to be treated, also Wilson lines have to be taken into account. First of all there are Wilson lines needed for ensuring gauge invariance of the correlators. Such Wilson lines come from diagrams with gluons coupling to partons involved in the hard scatterings. Furthermore, the presence of a soft factor is required, coupling the two correlators to each other. Factorization proofs ensure that the different contributions to the cross section factorize, see e.g.~Ref.~\cite{Diehl:2015bca} and references therein. The soft factor ensures a cancellation of the rapidity divergence and is required to define the subtracted DTMDs from the unsubtracted ones. We note that in the situation where all transverse distances are short (this is different from our situation), the soft factor for DPS has been calculated at the two loop level~\cite{Vladimirov:2016qkd}.

\begin{figure}[!tb]
\begin{center}
\includegraphics[width=0.250129\textwidth]{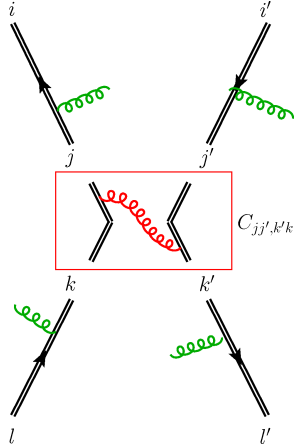}
\caption{Illustration of interactions in a single pair of two Wilson lines in the soft function. As explained in the main text, short-distance interactions (red online) are located in the boxed section in the center of this figure.}
\label{f:SPSsoft}
\end{center}
\end{figure}

We focus on a single pair of two Wilson lines first, which is illustrated in Fig.~\ref{f:SPSsoft}. Each double line in the Feynman graph represents a Wilson line given by
\begin{align}
W_{ij}(\bm{z},v) = \mathcal{P}\exp\left[
-igt^{a}\int_{-\infty}^{0}d\lambda\, v\,A^{\alpha}(z + \lambda v)\right]_{ij}^{z^{+} = z^{-} = 0}
\end{align}
and similarly for the adjoint representation. For the Wilson lines, the first argument indicates the position of the gauge field, whereas $v$ is a vector that is associated with the rapidity.

Consider a short-distance expansion of the Wilson line operator structure of the soft function that contains Wilson lines of the form $W(\bm{\xi}\pm\tfrac{1}{2}\bm{z},v)$ around $\bm{z}=0$. Short-distance interactions are then located closest to the point $\xi^{+} = \xi^{-} = 0$ of the Wilson lines, since this will result in a minimization of the number of Eikonal propagators with a large momentum. This is illustrated in Fig.~\ref{f:SPSsoft} for a single pair of two Wilson lines, where these kind of contributions are located in the boxed region in the center. It is necessary to keep open indices for the Wilson lines in this region, since this obect will appear in the matching procedure. A second pair of two Wilson lines should be included for DPS with the same type of short-distance interactions as the pair of two Wilson lines in Fig.~\ref{f:SPSsoft}. Although these short-distance interactions for the Wilson lines then are pairwise at the same point, the structure still contains nonperturbative interactions between Wilson lines separated by $\bm{y}$. In Fig.~\ref{f:SPSsoft} this corresponds to the gluons outside the box. Schematically, the structure of the soft function reads
\begin{align}
S(\bm{z}_1, \bm{z}_2, \bm{y}) & = C_{s}(\bm{z}_1)\, C_{s}(\bm{z}_2)\,S(\bm{y}), \label{e:soft-fact-match_A}
\end{align}
where $S(\bm{y})$ has half as many indices as $S(\bm{z}_1, \bm{z}_2, \bm{y})$, since it includes the open indices in the middle of the soft function. The coefficients $C_{s}$ are the contributions to the soft factor that contain all the $\bm{z}_{1}$ or $\bm{z}_{2}$ contributions respectively. 

\begin{figure}[!tb]
\begin{center}
\includegraphics[width=0.42\textwidth]{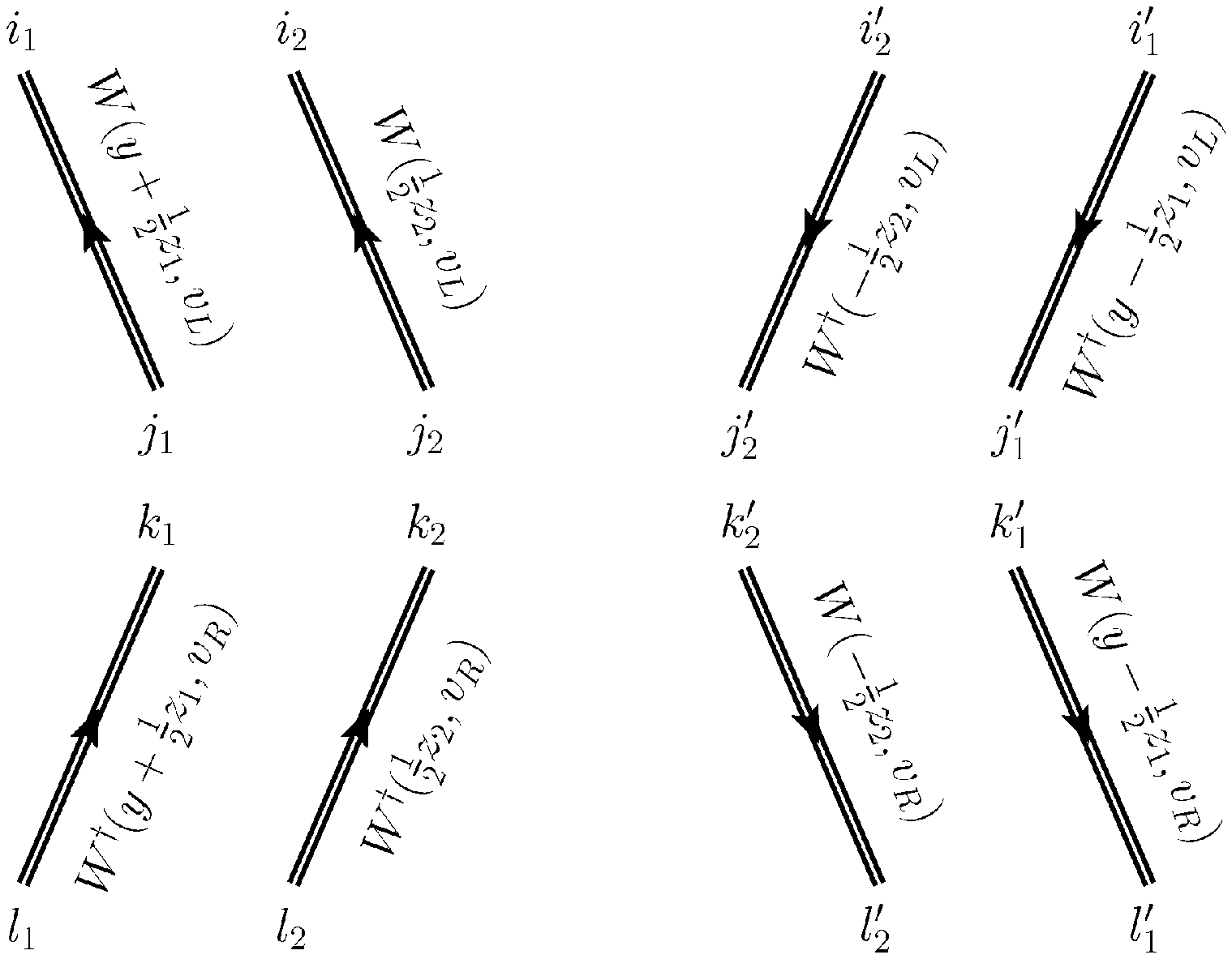}
\hspace{10mm}
\includegraphics[width=0.42\textwidth]{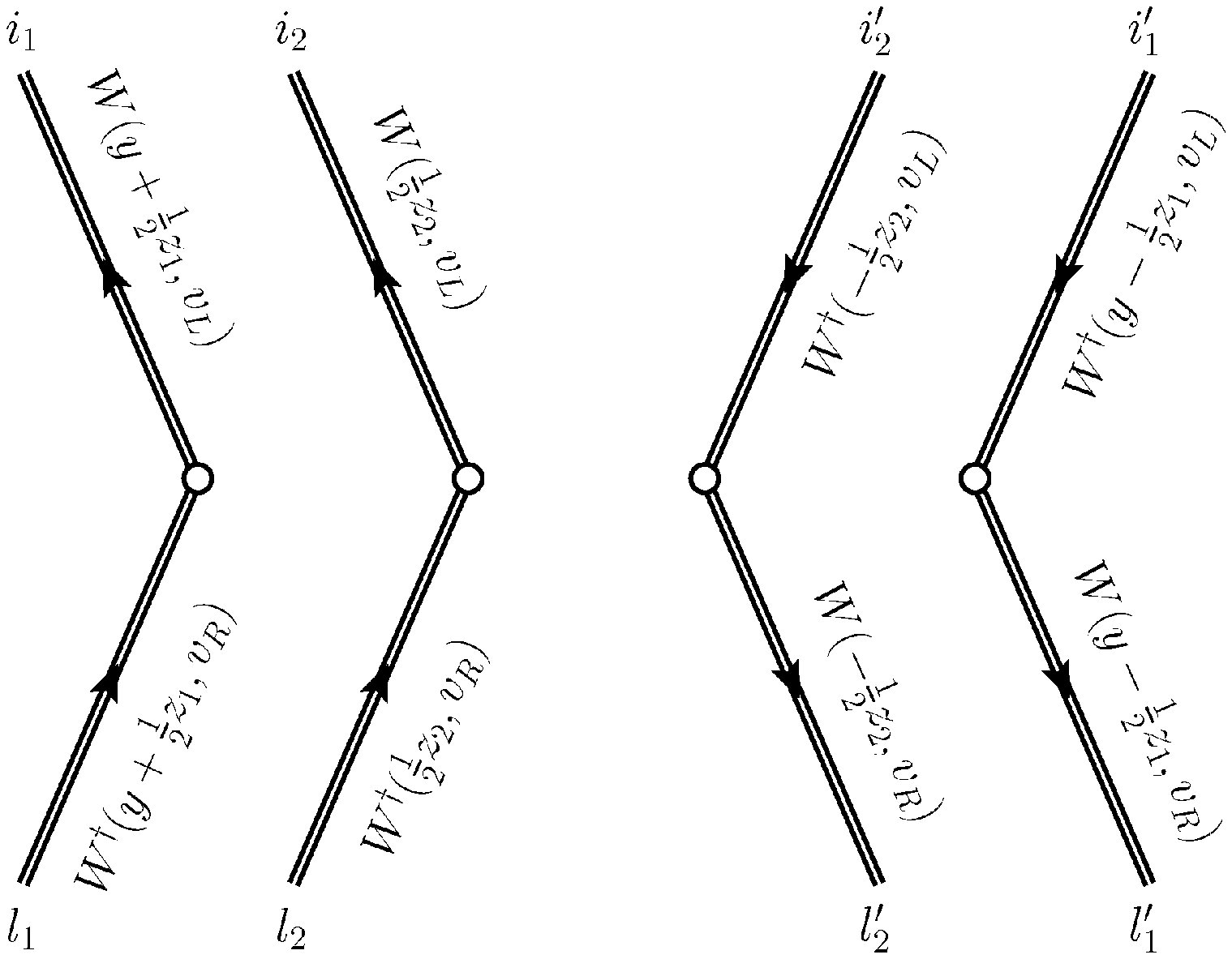} \\
(a) \hspace{65mm} (b)
\caption{(a) The soft function for double parton scattering decomposed in individual Wilson lines. Color projectors act on the indices in the middle. We have shown for $\bm{z}_{1} = \bm{z}_{2} = \bm{0}$ that it is equivalent to contract the indices in the middle and let the color projectors act on the indices at the ends of the Wilson lines, schematically giving us the structure in (b). Note that the picture includes gluon exchanges between the Wilson lines.}
\label{f:softstructure}
\end{center}
\end{figure}

The most general soft factor for DPS is illustrated in Fig.~\ref{f:softstructure}(a). Regarding the soft function in this figure, the indices $i$ at the top and the indices $l$ at the bottom represent the start and end of the Wilson lines at the two correlators, whereas the indices $j$ and $k$ in the middle involve the coupling of these Wilson lines. The indices are not contracted yet because we wish to do matching later on. Hard scattering in the DPS couples four parton lines, for which we choose to insert color projectors $P_{R}$ in order to simplify the color structure. The index $R$ labels the different color configuration that are possible (singlet, octet, etc.)~\cite{Kasemets:2014yna}, since the color structure for DPS has more possible configurations than for the SPS case. Two color projectors coupling the various indices in the middle of Fig.~\ref{f:softstructure}(a) are required, since there are two processes involved. Examples are the color singlet and octet quark projectors, given as
\begin{align}
P_{1}^{j_{1}j_{1}^{\prime}\,k_{1}k_{1}^{\prime}} = \frac{1}{N_c}\delta_{j_{1}j_{1}^{\prime}}\delta_{k_{1}k_{1}^{\prime}}, \hspace{15mm}
P_{8}^{j_{1}j_{1}^{\prime}\,k_{1}k_{1}^{\prime}} = 2 t_{j_{1}j_{1}^{\prime}}^{a}t_{k_{1}k_{1}^{\prime}}^{a}. \label{e:colorprojectors} 
\end{align}
For gluons more color projectors exist and mixed quark-gluon projectors also have to be considered~\cite{Diehl:2011yj,Kasemets:2014yna}, since in double parton scattering one of the partons could be a quark and the other one a gluon. We generalize the notation for the projectors introduced in Eq.~\ref{e:colorprojectors} as $P_{R}^{j_{i}j_{i}^{\prime}\,k_{i}k_{i}^{\prime}}$. In SPS the situation is simple and we would have $P_{1}^{jj^{\prime}}$ only.

Coming back to simplifying the soft function, we would in particular like to simplify the nonperturbative contribution $S(\bm y)$. Transforming the soft function in Fig.~\ref{f:softstructure}(a) to that in Fig.~\ref{f:softstructure}(b) is a convenient and simple way of ensuring this. For the nonperturbative sector, where it is desirable to have as few functions as possible, a simplification is especially helpful, since a procedure transforming the soft function in Fig.~\ref{f:softstructure}(a) to that in Fig.~\ref{f:softstructure}(b) reduces the number of open indices in the soft factor significantly. This transformation can be achieved by using color projectors for $S(\bm y)$ for $\bm{z}_{1} = \bm{z}_{2} = \bm{0}$. 

\begin{figure}[!tb]
\begin{center}
\includegraphics[width=0.42861896838\textwidth]{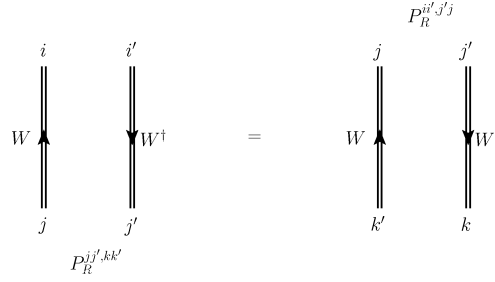}
\caption{Illustration of moving the color projector through the Wilson line structure. As explained before and illustrated in a previous figure, this allows for contracting the Wilson lines in the center region for $\bm{z}_{1} = \bm{z}_{2} = \bm{0}$. As before, note that the picture includes gluon exchanges between the Wilson lines.}
\label{f:softmatch}
\end{center}
\end{figure}

In order to prove that the Wilson lines can be contracted in a way that would allow the above sought simplification, we have to prove that we can commute color projectors acting on the Wilson lines in the soft factor through the Wilson lines. Then, the projectors would no longer be acting on the indices in the middle, but on indices at the ends of the Wilson lines and we could contract the Wilson lines, which would reduce the number of open indices. The identity we have to prove is
\begin{align}
W^{}_{ij}\, P_R^{jj',k'k}\, W^\dagger_{j'i'} & = W^{}_{jk}\, P_R^{ii',j'j}\, W^\dagger_{k'j'}, \label{e:comm-quark}
\end{align}
which is illustrated graphically in Fig.~\ref{f:softmatch}. We have proven this identity using the color Fierz identity
\begin{align}
2t_{ii^{\prime}}^{a}2t_{jj^{\prime}}^{a} & = \delta_{ij^{\prime}}\delta_{i^{\prime}j} - \frac{1}{N_c} \delta_{ii^{\prime}}\delta_{jj^{\prime}}.
\end{align}
Note that this relation only holds for the collinear situation, where the Wilson lines are at the same transverse position. For our purposes this is fine, since we will use it to study matching. The factorization also works for adjoint Wilson lines, which we need as soon as gluons are involved, implying
\begin{align}
W^{}_{ab}\, P_R^{bb',cc'}\, W^\dagger_{b'a'} & = W^{}_{bc}\, P_R^{aa',bb'}\, W^\dagger_{c'b'}. \label{e:comm-glu}
\end{align}

Using the color projector identity in Fig.~\ref{f:softmatch}, we can relate the full soft factor in Fig.~\ref{f:softstructure}(a) to the soft function in Fig.~\ref{f:softstructure}(b), where the indices in the middle are contracted. Rather than making the color projections at the color indices of the fields at $\xi^{+} = \xi^{-} = 0$, the same projection can be made for the indices of the fields at infinity. The fact that this relation holds implies that the collinear soft matrix for DPS is diagonal in the color representations of the left and right moving Wilson lines. The color structure of Eq.~\ref{e:soft-fact-match_A} then reads
\begin{align}
{}^{RR'}S_{a_1 a_2}(\bm{z}_1, \bm{z}_2, \bm{y}) & = {}^{R\,}C_{s, a_1}(\bm{z}_1)\, {}^{R\,}C_{s, a_2}(\bm{z}_2)\,{}^{RR}S(\bm{y})\, \delta_{RR'}^{} \,. \label{e:soft-fact-match_B}
\end{align}
Using projector identities one can also show that the soft factor ${}^{RR^{\prime}}S(\bm{y})$ is color diagonal. It follows from the proof of the above equation that the soft factor ${}^{RR'}S_{a_1 a_2}(\bm{z}_1, \bm{z}_1, \bm{y})$ is diagonal in the color representations $R$ and $R^{\prime}$ in the limit $|\bm{z}_1|,\,\,|\bm{z}_2|\,\,\ll\,\,|\bm{y}|$.

Now we come back to DPDs. They evolve with the evolution kernel ${}^{RR^{\prime}}K$, which is closely related to the soft factor ${}^{RR^{\prime}}S$ in a similar way as for SPS, but more complicated. The multiplicative structure of Eq.~\ref{e:soft-fact-match_B} results in
\begin{align}
{}^{RR'}K_{a_1 a_2}(\bm{z}_i,\bm{y}; \mu_i) & = \delta_{RR'}^{}\, \bigl[{}^{R}K_{a_1}(\bm{z}_1;\mu_1) + {}^{R}K_{a_2}(\bm{z}_2;\mu_2) + {}^{R}{J(\bm{y}; \mu_i)}\bigr], \label{e:CS-gen-match}
\end{align}
see a forthcoming paper for the details~\cite{Buffing:2016}. In Eq.~\ref{e:CS-gen-match} and all following equations, $\mu_{i}$ implies a dependence on both $\mu_1$ and $\mu_2$ and similarly for the other parameters where this notation is used. It should further be stressed that in contrast to SPS both the soft function ${}^{RR'}S$ and the evolution kernel ${}^{RR'}K$ are matrix valued expressions. The fact that the evolution kernel in Eq.~\ref{e:CS-gen-match} is a sum of three separate contributions simplifies dealing with evolution.

\section{Evolution equations}
\label{ss:evolution}
A description of DTMDs involves rapidity and scale parameters, for which evolution equations have to be derived and solved. Earlier descriptions of evolution for DTMDs using a different framework can be found in literature, see e.g.~the Refs.~\cite{Diehl:2011yj,Diehl:2011tt}. We give results for the situation $|\bm{z}_1|,\,\,|\bm{z}_2|\,\,\ll\,\,|\bm{y}|$, but as was the topic of the talk of M.~Diehl at this conference, many of our results have a wider applicability than the small distance expansion only. In these proceedings, on the other hand, we are looking at the small distance expansion, where the two hard processes are separated from each other. As such, we have two different renormalization scales, $\mu_{1}$ and $\mu_{2}$, for which we need two separate evolution equations, namely
\begin{align}
\frac{\partial}{\partial \log\mu_1}\,{}^{R}F_{a_1 a_2}(x_i,\bm{z}_i,\bm{y};\mu_i,\zeta) & = \gamma_{F, a_1}(\mu_1, x_1\zeta/x_2)\,{}^{R}F_{a_1 a_2}(x_i,\bm{z}_i,\bm{y};\mu_i,\zeta) \label{e:RG-TMD}
\end{align}
and a similar equation for the $\mu_2$ evolution. Furthermore, the DTMD carries a color representation index $R$. Here, the $\gamma_{F,a_i}$ are anomalous dimensions of the DTMDs, equal to the same objects in the TMD evolution~\cite{Collins:2011zzd,Aybat:2011zv}. Also, $\gamma_{F,a}$ depends only on whether one is dealing with (anti)quarks, involving fundamental Wilson lines, or with gluons, involving adjoint Wilson lines. In the $\mu$-evolution equations, the anomalous dimensions $\gamma_{F}$ depend on the rapidity regularization scale $\zeta$. A closer analysis shows that $\zeta$ has to be rescaled with $x_{1}/x_{2}$ or $x_{2}/x_{1}$. This rescaling is required, since we have chosen to use a single $\zeta$ scale per DTMD. It then follows from Eq.~\ref{e:RG-TMD} that the $\mu$ evolution of DTMDs is given by
\begin{align}
{}^{R}F_{a_1 a_2}(x_i,\bm{z}_i,\bm{y};\mu_i,\zeta) = & {}^{R}F_{a_1 a_2}(x_i,\bm{z}_i,\bm{y};\mu_{0i},\zeta) \nonumber \\[0.2em]
& \qquad \times\exp\biggl[ \int_{\mu_{01}}^{\mu_1} \frac{d\mu}{\mu}\,\gamma_{F,a_1}(\mu, x_1\zeta/x_2) + \int_{\mu_{02}}^{\mu_2} \frac{d\mu}{\mu}\,\gamma_{F,a_2}(\mu, x_2\zeta/x_1) \biggr] \label{e:RG-TMD-sol}
\end{align}
from the starting scales $\mu_{01}$ and $\mu_{02}$ for $\mu_{1}$ and $\mu_{2}$. The rapidity evolution of DTMDs is given by
\begin{align}
\frac{\partial}{\partial\log \zeta} 
{}^{R}{F_{a_1 a_2}}(x_i,\bm{z}_i,\bm{y}, \mu_i,\zeta) & = \frac{1}{2} \sum_{R'}{}^{RR'}{K_{a_1 a_2}(\bm{z}_i,\bm{y}; \mu_i)}\,{}^{R'}{F_{a_1 a_2}(x_i,\bm{z}_i,\bm{y}; \mu_i,\zeta)} \, , \label{e:CS-TMD}
\end{align}
where we can split the evolution kernel $K$ in the short-distance limit into the three separate contributions ${}^{R}K_{a_1}(\bm{z}_1;\mu_1)$, ${}^{R}K_{a_2}(\bm{z}_2;\mu_2)$ and ${}^{R}{J(\bm{y}; \mu_i)}$ as in Eq.~\ref{e:CS-gen-match}.

The anomalous dimensions $\gamma_{F}$ in the $\mu$ evolution equations contain a $\zeta$ dependence and the evolution kernel $K$ in the $\zeta$ evolution equation is $\mu$-dependent. Before we can write down the full DTMD evolution equation, it has to be understood how they evolve. The evolution kernel ${}^{RR^{\prime}}K$ can be written as a sum of terms as in Eq.~\ref{e:CS-gen-match}. Taking the derivative with respect to $\mu_1$ of the separate contributions gives
\begin{align}
\frac{\partial}{\partial \log\mu_1}\, {}^{R} K_a(\bm{z}; \mu_1) & = - {}^{R}\gamma_{K,a}(\mu_1), \label{e:CS-coll-RG_K} \\
\frac{\partial}{\partial \log\mu_1}\, {}^{R}J(\bm{y}; \mu_i) &= - {}^{R\,}\gamma_J(\mu_1) \label{e:CS-coll-RG_J}
\end{align}
and similar equations for the derivative with respect to $\mu_2$. The anomalous dimensions in the Eqs.~\ref{e:CS-coll-RG_K} and \ref{e:CS-coll-RG_J} satisfy
\begin{align}
\gamma_{K,a}(\mu) &= {}^{R}\gamma_{K,a}(\mu) + {}^{R}\gamma_J (\mu) . \label{e:AD-sum}
\end{align}
Furthermore, the rapidity dependence of the anomalous dimension $\gamma_{F,a_1}$ that came from the $\mu$-scale equation is given by
\begin{align}
\frac{\partial}{\partial \log\zeta}\, \gamma_{F,a}(\mu, \zeta) &= - \frac{1}{2}\mskip 1.5mu \gamma_{K,a}(\mu) \, . \label{e:cusp}
\end{align}
Note that the $\mu$ dependence of the anomalous dimensions is understood to be through the coupling $\alpha_{s}(\mu)$ only. Combining the above information regarding the evolution with respect to $\mu_1$, $\mu_2$ and $\zeta$, the solution of the evolution equations for DTMDs is then given by
\begin{align}
{}^{R}F_{a_1 a_2}(x_i,\bm{z}_i,\bm{y};\mu_i,\zeta) & = {}^{R}F_{a_1 a_2}(x_i,\bm{z}_i,\bm{y};\mu_{0i},\zeta_0) \nonumber \\
& \quad \times\exp\,\biggl\{ \int_{\mu_{01}}^{\mu_1} \frac{d\mu}{\mu}\,\biggl[\gamma_{F,a_1}(\mu, \mu^2) - \gamma_{K,a_1}(\mu) \log\frac{\sqrt{x_1\zeta/x_2}}{\mu} \biggr] \nonumber \\
& \qquad\qquad + \int_{\mu_{02}}^{\mu_2} \frac{d\mu}{\mu}\,\biggl[ \gamma_{F,a_2}(\mu, \mu^2) - \gamma_{K,a_2}(\mu) \log\frac{\sqrt{x_2\zeta/x_1}}{\mu} \biggr] \nonumber \\
& \qquad\qquad + \Bigl[ {}^{R}K_{a_1}(\bm{z}_1,\mu_{01}) + {}^{R}K_{a_2}(\bm{z}_2,\mu_{02}) + {}^{R}J(\bm{y},\mu_{0i}) \Bigr]\log\frac{\sqrt{\zeta}}{\sqrt{\zeta_0}} \biggr\} \label{e:evsolving}
\end{align}
for the starting scales $\mu_{01}$, $\mu_{02}$ and $\zeta_{0}$.

\section{Matching}
\label{ss:matching}
The matching equation for DTMD/DPDF matching is given by~\cite{Buffing:2016}
\begin{align}
& {}^{R}F_{a_1 a_2}(x_i,\bm{z}_i,\bm{y};\mu_i,\zeta) & = \sum_{b_1 b_2} 
	{}^{R\,}C\!_{a_1 b_1}(x_1',\bm{z}_1;\mu_{1},\mu_{1}^2) \underset{x_1}{\otimes}
	{}^{R\,}C\!_{a_2 b_2}(x_2',\bm{z}_2;\mu_{2},\mu_{2}^2) \underset{x_2}{\otimes}
	{}^{R}F_{b_1 b_2}(x_i',\bm{y};\mu_{i},\zeta), \label{e:matching}
\end{align}
where the convolution between two functions $A$ and $B$ is given by
\begin{align}
A(x') \underset{x}{\otimes} B(x') &= \int_{x}^1 \frac{dx'}{x'}\, A(x')\,B\biggl(\frac{x}{x'}\biggr). \label{e:conv-def}
\end{align}
The summation over $b_{1}$ and $b_{2}$ is over parton species and polarization as in Ref.~\cite{Diehl:2011yj}. The two coefficient functions are both TMD/PDF matching coefficient functions, a statement that can be seen more easily when writing down the formalism for DTMD/DPDF matching at the level of operators~\cite{Buffing:2016}.

In Eq.~\ref{e:evsolving} the evolution of DTMDs from a set of starting scales is given. Combining this result with the matching equation in Eq.~\ref{e:matching}, we can write the matching equation for the DTMD in the short-distance limit, where the rapidity dependence of the coefficient functions will be split off in separate terms containing ${}^{R}K\!_{a_1}(\bm{z}_1,\mu_{01})$ and ${}^{R}K\!_{a_2}(\bm{z}_2,\mu_{02})$. It can be shown that this transforms the matching equation into
\begin{align}
& {}^{R}F_{a_1 a_2}(x_i,\bm{z}_i,\bm{y};\mu_i,\zeta)
\nonumber \\
&\quad = \sum_{b_1 b_2} 
	{}^{R\,}C\!_{a_1 b_1}(x_1',\bm{z}_1;\mu_{01},\mu_{01}^2) \underset{x_1}{\otimes}
	{}^{R\,}C\!_{a_2 b_2}(x_2',\bm{z}_2;\mu_{02},\mu_{02}^2) \underset{x_2}{\otimes}
	{}^{R}F_{b_1 b_2}(x_i',\bm{y};\mu_{0i},\zeta_0)
\nonumber \\
& \qquad \times \exp\, \biggl\{ 
		\int_{\mu_{01}}^{\mu_1} \frac{d\mu}{\mu}\,
	\biggl[ \gamma_{F,a_1}(\mu, \mu^2) - \gamma_{K,a_1}(\mu) \log\frac{\sqrt{x_1\zeta/x_2}}{\mu} \biggr]
		+ {}^{R}K\!_{a_1}(\bm{z}_1,\mu_{01}) \log\frac{\sqrt{x_1\zeta/x_2}}{\mu_{01}}
\nonumber \\
& \qquad\qquad\;\, + \int_{\mu_{02}}^{\mu_2} \frac{d\mu}{\mu}\,
	\biggl[ \gamma_{F,a_2}(\mu, \mu^2) - \gamma_{K,a_2}(\mu) \log\frac{\sqrt{x_2\zeta/x_1}}{\mu} \biggr]
		+ {}^{R}K\!_{a_2}(\bm{z}_2,\mu_{02}) \log\frac{\sqrt{x_2\zeta/x_1}}{\mu_{02}}
\nonumber \\
& \qquad\qquad\;\, + {}^{R}J\!(\bm{y},\mu_{0i}) \log\frac{\sqrt{\zeta}}{\sqrt{\zeta_0}} \biggr\}. \label{e:small-z-evolved}
\end{align}
For color singlet configurations the above combined evolution and matching equation in essence consists of doubling the single TMD/PDF formalism. For other color configurations there is an additional Sudakov suppression coming from ${}^{R}J\!(\bm{y},\mu_{0i})$, which is zero for color singlet configurations. Note that the evolution kernel contributions ${}^{R}K\!_{a_1}(\bm{z}_1,\mu_{01})$ and ${}^{R}K\!_{a_2}(\bm{z}_2,\mu_{02})$ have a color dependence.

At the level of the cross section for DPS in proton-proton collisions, two DTMDs have to be involved. The matching equation for a cross section contribution involving two such objects is given by
\begin{align}
W_{\text{large $\bm{y}$}} &= \sum_{c_1 c_2 d_1 d_2} \sum_{R} \exp\, \biggl\{ 
			\int_{\mu_{01}}^{\mu_1} \frac{d\mu}{\mu}\,
	\biggl[\gamma_{F,a_1}(\mu, \mu^2)
		- \gamma_{K,a_1}(\mu) \log\frac{{Q_1^2}}{\mu^2} \biggr]
		+ {}^{R}K\!_{a_1}(\bm{z}_1,\mu_{01})
			\log\frac{{Q_1^2}}{\mu_{01}^2} \nonumber \\
& \qquad\qquad + \int_{\mu_{02}}^{\mu_2} \frac{d\mu}{\mu}\,
	\biggl[ \gamma_{F,a_2}(\mu, \mu^2)
		- \gamma_{K,a_2}(\mu) \log\frac{{Q_2^2}}{\mu^2} \biggr]
		+ {}^{R}K\!_{a_2}(\bm{z}_2,\mu_{02}) \log\frac{{Q_2^2}}{\mu_{02}^2} \biggr\} \nonumber \\[0.3em]
& \quad \times
	{}^{R\,}C\!_{b_1 d_1}(\bar{x}_1',\bm{z}_1;\mu_{01},\mu_{01}^2) \underset{\bar{x}_1}{\otimes}
	{}^{R\,}C\!_{b_2 d_2}(\bar{x}_2',\bm{z}_2;\mu_{02},\mu_{02}^2) \,\underset{\bar{x}_2}{\otimes} \nonumber \\[0.3em]
& \quad \times
	{}^{R\,}C\!_{a_1 c_1}(x_1',\bm{z}_1;\mu_{01},\mu_{01}^2) \underset{x_1}{\otimes}
	{}^{R\,}C\!_{a_2 c_2}(x_2',\bm{z}_2;\mu_{02},\mu_{02}^2)\,\underset{x_2}{\otimes} \nonumber \\
& \quad \times \bigl[ \Phi(\nu\bm{y}) \bigr]^2\,
	\exp\,\biggl[{}^{R}J\!(\bm{y},\mu_{0i})\log\frac{\sqrt{Q_1^2 \mskip 1.5mu Q_2^2}}{\zeta_0} \,\biggr]
	{}^{R}F_{d_1 d_2}(\bar{x}_i,\bm{y};\mu_{0i},\zeta_0)\,
	{}^{R}F_{c_1 c_2}(x_i,\bm{y};\mu_{0i},\zeta_0). \label{e:W-large-y}
\end{align}
In this equation, the energies $Q_{1}$ and $Q_2$ of the two hard partonic interactions are related to the scale parameters $\zeta$ and $\overline{\zeta}$ of the two DTMDs through the relation $\zeta\overline{\zeta} = Q_{1}^{2}Q_{2}^{2}$. Coming from Eq.~\ref{e:small-z-evolved}, each DTMD contributes two coefficient functions, giving a grand total of four for the cross section contribution. In addition, there is a $\bm{y}$ contribution in the form of $\bigl[ \Phi(\nu\bm{y}) \bigr]^2$. This function regulates the ultraviolet region and ensures that the integral converges at small distances~\cite{Diehl:2016khr}. It should be noted that the Eqs.~\ref{e:small-z-evolved} and \ref{e:W-large-y} are valid for both quarks and gluons.

\section{Discussions and conclusions}
\label{ss:conclusions}
In our work we use the short-distance expansion, valid if the two partons initiating the two different hard processes in DPS have perturbative transverse momenta $\bm{k}_1$ and $\bm{k}_2$. We also consider the large $\bm{y}$ situation, such that the hard processes are spatially well separated from each other. In this limit the DPS soft function can be factorized in three separate contributions, namely $\bm{z}_1$, $\bm{z}_2$ and $\bm{y}$-dependent ones. From this, it follows that the evolution kernel has three separate terms. Using this important result we have given the evolution equations for DPDs and solved them. We furthermore presented the matching equations for DTMDs as well as for the cross section contribution for the production of colorless final states.

An important result is that the matching equation for the DTMDs, Eq.~\ref{e:small-z-evolved}, has two coefficient functions, with each of them equal to a single TMD/PDF coefficient function. The reason for this is the use of the short-distance expansion and considering the large $\bm{y}$ situation. As such, for DPS the coefficient functions can be recycled from the coefficient functions for the TMD/PDF matching. In a derivation at the level of operators this is apparent from the start. In our forthcoming paper~\cite{Buffing:2016} we will give the matching coefficients for all polarization modes, complementing results for TMDs in SPS~\cite{Collins:2011zzd,Aybat:2011zv,Bacchetta:2013pqa,Echevarria:2015uaa}.

\acknowledgments
TK is supported by the European Community under the ``Ideas'' program QWORK (contract 320389).



\end{document}